# Influence of complete energy sorting on the characteristics of the odd-even effect in fission-fragment element distributions

Beatriz Jurado and Karl-Heinz Schmidt

*CENBG, CNRS/IN2P3, Chemin du Solarium B. P. 120, 33175 Gradignan, France*

**Abstract.** The characteristics of the odd-even effect in fission-fragment Z distributions are compared to a model based on statistical mechanics. Special care is taken for using a consistent description for the influence of pairing correlations on the nuclear level density. The variation of the odd-even effect with the mass of the fissioning nucleus and with fission asymmetry is explained by the important statistical weight of configurations where the light nascent fission fragment populates the lowest energy state of an even-even nucleus. This implies that entropy drives excitation energy and unpaired nucleons predominantly to the heavy fragment. Therefore, within our model, the odd-even effect appears as an additional signature of the recently discovered energy-sorting process in nuclear fission.

1. Introduction

Nuclear fission is a large-scale collective motion involving a drastic rearrangement of the nucleons present in the initial fissioning system. The evolution from a mononuclear to a di-nuclear system results from the interplay of different structural and dynamical effects. Several theoretical models have shown that the di-nuclear configuration, in which the fragments have acquired their individual properties, takes over well before scission (see section 4.1). Therefore, nuclear fission offers a unique possibility to investigate the behaviour of two nuclei in contact with low relative velocities.

In fission, there is a huge number of possible states near scission involving hundreds of different fragments that follow continuous distributions for different deformations, kinetic energies, excitation energies and angular momenta. Therefore, statistical mechanics appears as an appropriate tool to investigate the fission process and to identify the main mechanisms that are responsible for the evolution of the fissioning system. In [1-3] we used statistical mechanics to investigate the partition of excitation energy between the nascent fragments in low-energy fission. This study resulted in the discovery of the energy-sorting process, a peculiar mechanism where most of the excitation energy that is available before scission is transferred to the heavy fragment [1]. The energy sorting is clearly reflected by the average number of prompt neutrons emitted by the fragments. Indeed, when the available excitation energy is increased, the number of prompt neutrons emitted by the heavy fragment increases, while the number of prompt neutrons emitted by the light fragment remains unchanged [1].

In this work, we use statistical mechanics to investigate the enhanced production of fission fragments with an even number of protons, i.e. the so-called odd-even effect. Statistical arguments have been used previously to explain in a more or less quantitative way different aspects of the odd-even effect in fission-fragment yields (see chapter 3). However, a comprehensive model that explains the odd-even staggering in all its complexity is not yet available. Here, we apply statistical mechanics in the most complete way by considering equilibration in all the degrees of freedom relevant to the description of the odd-even effect. These degrees of freedom are the number of protons, the number of neutrons and the excitation energies of the two nascent fragments. We will show that, within this model, complete energy sorting, i.e. the process in which the light fragment is frozen and populates the states below the pairing gap of an even-even nucleus, is the main mechanism responsible for the odd-even effect in fission-fragment charge distributions.

## 2. Experimental systematics of the odd-even effect

Fig. 1 shows the measured fission-fragment yields as a function of proton number for the fissioning nucleus $^{229}$Th, which was excited in the Coulomb field of lead target atoms slightly above the fission barrier [4]. The average excitation energy of the fissioning nucleus is 11 MeV, corresponding to the excitation energy induced by neutrons with 5.8 MeV kinetic energy. This experiment allowed measuring the odd-even structure continuously over a large range of mass splits. This was not possible in heavier actinides due to the extremely low yields for symmetric splits. The global shape of the data from Fig. 1 can be described by three humps, one centered at symmetry ($Z\approx45$) and two at asymmetry ($Z\approx36$ and 54). These humps result from the shape of the potential as a function of mass (or charge) asymmetry as given by the liquid-drop model with the influence of shell effects [5]. The odd-even effect in fission-fragment yields is the fine structure that is superimposed to the gross shape of the yields showing an enhanced production of fragments with even $Z$.

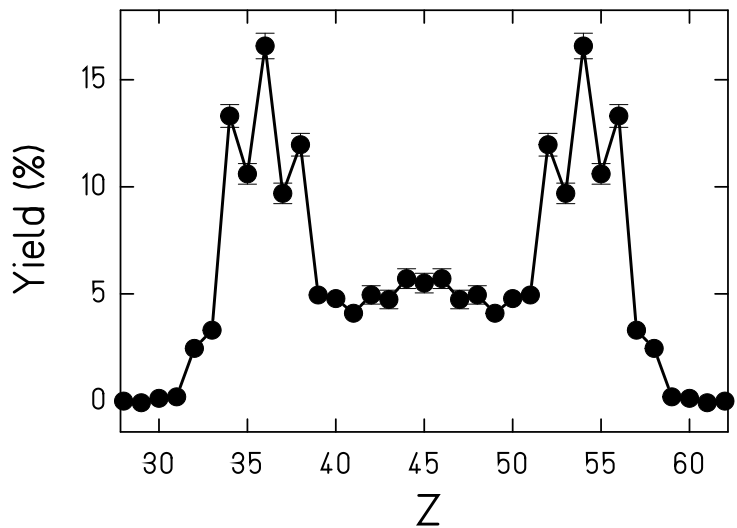

**Figure 1.** Element distribution observed in the electromagnetic-induced fission of $^{229}$Th [4].

It was proposed in [6] to quantify the odd-even staggering by the local odd-even effect $\delta_p(Z)$, which corresponds to the third differences of the logarithm of the yields $Y$:

$$\delta_p(Z) = \frac{1}{8}(-1)^{Z_{low}}\left[\ln Y(Z_{low}+3) - \ln Y(Z_{low}) - 3(\ln Y(Z_{low}+2) - \ln Y(Z_{low}+1))\right] \quad (1)$$

where $Z_{low} = Z - 3/2$. The quantity $\delta_p(Z)$ filters out from the yields the variations that extend over a significant number of charges and are related to the global shape of the potential energy. We can consider two curves; one links the logarithms of the yields of neighbouring even-$Z$ fragments $\ln Y_{Z=e}(Z)$, and the other connects the logarithms of the yields of neighbouring odd-$Z$ fragments $\ln Y_{Z=o}(Z)$. These two curves are continuous functions (if the yields $Y_{Z=e}$ and $Y_{Z=o}$ follow a Gaussian shape, the curves are parabolas) and can be evaluated for any value of $Z$. As explained in [6], $\delta_p(Z)$ equals half the distance between the two curves:

$$\delta_p(Z) = 0.5(\ln Y_{Z=e}(Z) - \ln Y_{Z=o}(Z)) \quad (2)$$

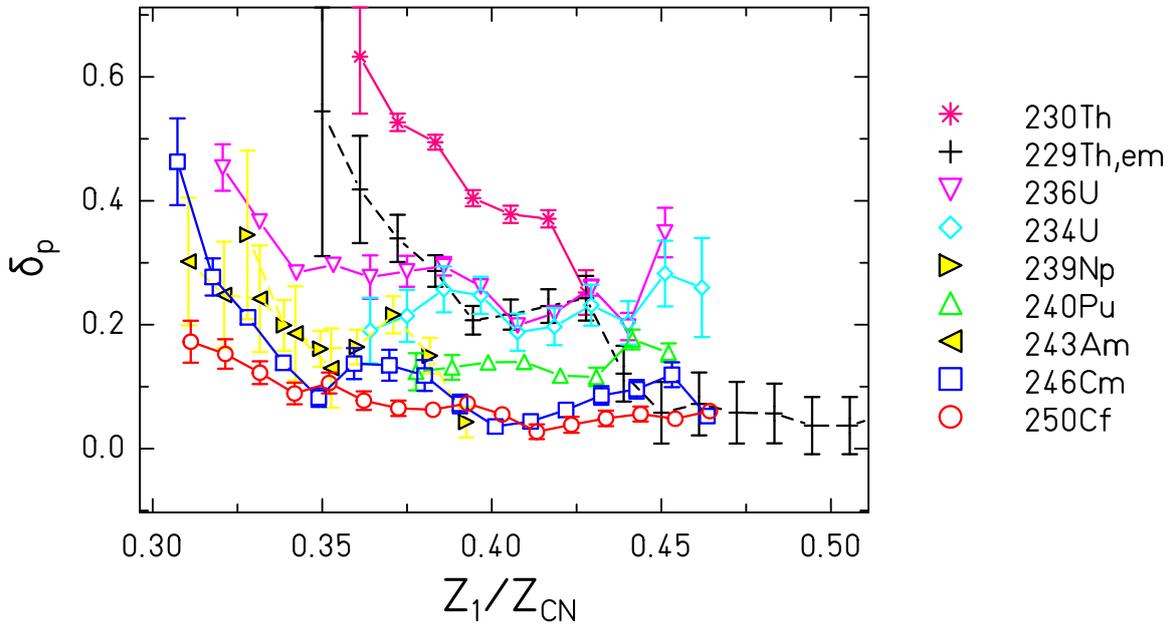

**Figure 2:** (Colour online) Measured local odd-even effect $\delta_p$ as a function of asymmetry. The legend indicates the fissioning nucleus.

Fig. 2 shows experimental results on $\delta_p(Z)$ as a function of charge asymmetry for different fissioning nuclei. In this figure, the charge asymmetry has been parameterized as the ratio of the charge of the light fragment $Z_1$ and the charge of the fissioning nucleus $Z_{CN}$. The curve named $^{229}$Th,em has been obtained from the data shown in Fig. 1. The remaining data, corresponding to thermal-neutron-induced fission, have been taken from the compilation given in [7]. Fig. 2 illustrates several general trends:

(i) The amplitude of $\delta_p$ decreases with increasing mass of the fissioning system and with excitation energy (cf. $^{229}$Th,e.m.).
(ii) For a given fissioning nucleus $\delta_p$ increases with asymmetry.

(iii) Also odd-Z fissioning systems like $^{243}$Am or $^{239}$Np show an odd-even effect at large asymmetry whose magnitude is about the same as for even-Z systems of comparable mass.

### 3. Previous explanations for the odd-even effect

The theoretical interpretation of the odd-even effect in fission-fragment yields was inspired for a long time by the observation that the magnitude of the effect is very sensitive to the initial excitation energy of the fissioning system and that no odd-even effect had been observed in odd-Z fissioning systems. Thus, the odd-even effect seemed to be a measure for the survival of a completely paired proton configuration at scission [8]. Based on statistical concepts, several authors attempted to relate the magnitude of the odd-even structure in the $Z$ yields with the intrinsic excitation energy available in the fissioning system in the vicinity of the scission point [9, 10]. These models had two key parameters: the probability $q$ to break a proton pair when the required energy is available and the probability $p$ for the protons of a broken pair to end up in the same fragment. They were not based on the number of available states of the system that is the key quantity of statistical mechanics. Moreover, in these models the even-odd effect is expected to vanish for odd-Z fissioning nuclei [8], which is in clear contradiction with the experimental observation (see Fig. 2).

Some attempts were made to theoretically study the dynamical process of pair breaking in the fission process. The analysis of several fission observables performed in [10] and [11] lead the authors to the conclusion that most of the nucleonic excitations take place during the sudden snapping of the neck. According to [10, 11], the decrease of the odd-even staggering with increasing mass of the fissioning nucleus is explained by the dynamics of the neck rupture. The probability of breaking a pair is directly related to the velocity of the neck rupture, which is shown to increase with the Coulomb repulsion at the scission point. This leads to the generation of more excitation energy for heavy than for lighter fissioning nuclei. However, this work did not consider the odd-even effect of odd-Z fissioning nuclei and the strong asymmetry-dependence of the odd-even effect.

Qualitative arguments for the increase of the odd-even effect with asymmetry of odd-Z fissioning nuclei were given on the basis of the mass dependence of the pairing gap and of the single-particle level density in [12]. In [12] and in [13], a simple model calculation was presented where the probability of the unpaired proton to be attached to one of the two forming fragments was assumed to be proportional to the number of protons of the fragments. It was also assumed that no proton pairs were broken on the way from saddle to scission. However, this calculation does not agree with the experimental data.

The scope of the work by Steinhäuser et al. [14] was to demonstrate that the manifestation of an odd-even effect for odd-Z fissioning systems at large mass asymmetry is well understandable in the framework of statistical mechanics. The model by Steinhäuser et al. explains the observed increase of the odd-even effect with asymmetry by the assumption that unpaired protons stick to the fragments in proportion their single-particle level density. A restriction of this model is that it considers a fixed probability distribution for the number of unpaired protons. According to Steinhäuser et al., the odd-even effect observed for even-even fissioning nuclei at symmetry is due to the existence of configurations where the excitation energy is only used to break neutron pairs. The probability for these configurations in the mononuclear regime has been calculated by Rejmund et al. on the basis of the superfluid model using statistical considerations [15]. Rejmund et al. did not consider the emergence of the nascent fragments with their individual properties.

Very recently, Möller et al. [16] have calculated the available energies of the fissioning nucleus for symmetric and asymmetric configurations at the point where the nascent fragments have nearly established their final properties. The fission trajectories were determined on the basis of overdamped random walks on a five dimensional potential-energy surface. This work shows a dependence of the available energy with mass asymmetry, which, according to the authors, correlates well with the odd-even staggering in fission yields. However, only qualitative arguments are given.

### 4. Principle of the model

We present a model for the even-odd effect that goes well beyond the models by Steinhäuser et al. [14] and F. Rejmund et al. [15]. It can be resumed as the equilibration of two nuclei in contact that share a finite total amount of excitation energy and follow level densities in which the effects of pairing correlations are consistently included. The equilibration is ruled by the physical quantities of relevance in statistical mechanics: the total available excitation energy and the density of states of the two fragments. Note that statistical equilibrium represents the asymptotic state to which any dynamical process is driven to. Therefore, our approach will serve to reveal the essential physical process behind the even-odd effect in fission-fragment elemental yields within the frame of statistical mechanics.

A possible variation of the available excitation energy, e.g. as a function of the asymmetry of the mass split that is stressed in ref. [16], is beyond the scope of this work. This effect can easily be included if the necessary information is available from some reliable theoretical estimation.

### 4.1. Definition of the starting point

Because the number of microstates has a minimum on the fission path close to the outer fission barrier, the excited nucleus can only pass the barrier with a sizeable probability if the different types of degrees of freedom (intrinsic and collective) are in statistical equilibrium [17]. This implies that the available energy above the fission barrier, which is equal to the total energy of the nucleus minus the height of the outer fission barrier, is equally shared between the different degrees of freedom. Most of the available states are intrinsic excitations. Thus, most of the excitation energy available above the outer saddle is stored in intrinsic excitations.

The intrinsic excitation energy grows on the way from saddle to scission because part of the potential-energy release is dissipated into intrinsic excitations. Different mechanisms lead to the transformation of collective energy into intrinsic excitation energy during the descent from saddle to scission: one-body [18] and two-body dissipation due to level crossing [19]. For simplicity, we assume that all the dissipated energy is generated early enough, and that the fission process is slow enough for the fragments to reach equilibrium in the population of their intrinsic states before scission. The dissipated energy increases with the mass of the fissioning nucleus since the fission barrier is located at smaller deformations, and the saddle-to-scission path is extended [20]. Note that we do not consider the additional intrinsic excitation that may appear during the neck rupture.

Theoretical investigations of the gradual transition from the mononucleus regime to the dinuclear system concerning shell effects [21, 22], pairing correlations [23] and congruence energy [24] show that the fission-fragment properties are already rather well established in the vicinity of the outer saddle. As explained in [21], the strong influence of the fragment shells on the single-particle states well before scission is due to a fundamental quantum-mechanical effect in which the nucleons are localized in the two fragments as soon as there is some necking. Very recently, Hartree-Fock calculations with BCS pairing residual interaction have been performed for $^{264}$Fm in [18]. These calculations confirm the very early onset of the fission fragment properties. One can see that the establishment of the shell gaps $Z = 50$ and $N = 82$ of the $^{132}$Sn nascent fragments and the vanishing of the proton and neutron pairing energies occur well before the configuration where the neck disappears. Therefore, it is reasonable to assume that well before scission the fissioning system consists of two well-defined nuclei in contact through the neck. At this point the total amount of excitation energy $U_{tot}$ available is equal to the sum of the intrinsic excitation energy above the outer saddle and the energy acquired by dissipation from saddle to scission. Intrinsic excitations are expected to be homogeneously distributed within the nuclear volume. This is likely to hold also in the transition from a mononuclear to a dinuclear system that takes place very rapidly near the outer saddle [21]. Consequently, a reasonable assumption is that $U_{tot}$ is initially shared among the fragments according to the ratio of their masses.

### 4.2. Equilibration of two nuclei in contact

We assume that the system formed by the two nuclei in contact then evolves to a state of statistical equilibrium, the macrostate of maximum entropy, where all the available microstates have equal probability [25]. This implies that the total available energy will be distributed among the two nascent fragments according to the probability distribution of the available microstates which is given by the total nuclear level density[*]. Eventually, energy sorting will take place, and the light fragment will transfer essentially all its excitation energy to the heavy one [1, 3]. Nucleon exchange between the fragments leads also to an equilibration between even-even, even-odd, odd-even, and odd-odd light and heavy nascent fragments with a restriction on the total energy and on the gross mass asymmetry given by the bottom of the potential in the fission valleys.

---

[*] The degeneracy of magnetic substates is not considered because it contributes very little to the variation of the state density with excitation energy.

To obtain the probability of populating a given configuration at statistical equilibrium and derive the local odd-even effect $\delta_p$, we have to consider the level densities of neighboring even-even, even-odd, odd-even or odd-odd nuclei in an absolute energy scale. However, it is not appropriate to use the binding energies of the fragments as the origin of the level densities because this would lead to the determination of the fragment yields, including the slowly-varying components that are filtered out by the quantity $\delta_p$. Therefore, we need a scale for the level densities where these effects are filtered out as well. As illustrated when discussing the meaning of $\delta_p$, we can consider a smooth surface (in the neutron number $N$ and $Z$ space) that connects the yields of odd-odd fragments. This surface can be associated with the potential energy in the fission valley for the formation of odd-odd pre-fragments in the case of an even-even fissioning nucleus. The potential surface of odd-mass pre-fragments is at an energy -Δ (Δ being the corresponding pairing gap) with respect to the surface of odd-odd pre-fragments, and the potential surface of even-even pre-fragments is at an energy of -2Δ. Thus, the required filtering of the level densities can be obtained by placing the level densities in a reduced energy scale $U = E_{gs} - n\Delta$[†] where the excitation energies above the ground state $E_{gs}$ of even-even nuclei are lowered by 2Δ ($n = 2$), those of odd-mass nuclei by Δ ($n = 1$) and are left unchanged for odd-odd nuclei ($n = 0$).

On Fig. 3, we have applied the reduced energy scale to experimental level densities determined by the Oslo method [26] of various nuclei located around $A$=165 and 45 using $\Delta = 12/\sqrt{A}$. In this work we consider that these experimental level densities, which are derived for states at ground-state deformation, are also appropriate for states beyond the saddle point (the pertinence of this approximation is discussed in section 6). Therefore, these two groups of experimental level densities represent the level densities of two deformed, touching fission fragments corresponding to a very asymmetric split. Within the heavy group, the level densities of neighbouring even-even and even-odd nuclei are almost identical. Sizeable differences appear only in the energy interval $-2\Delta_2 < U_2 < -\Delta_2$, where only even-even nuclei have states. For the light-mass group, the level densities converge well at positive reduced energies. Some fluctuations are present at negative reduced energies due to the melting of Cooper pairs. Fig. 3 clearly shows that the logarithmic slope of the level densities is nearly constant and that the logarithmic slope of the heavy group is much larger than the one of the light group.

The total amount of possible configurations (or microstates) with particular values of $Z_1$ and $Z_2$ is directly related to the integral of the total level density for that particular split over the excitation energy of one fragment $U_i$, with the condition that $U_1+U_2=U_{tot}$. This integral reflects the freedom of the system in the division of excitation energy. Note that $U_1$, $U_2$, and $U_{tot}$ are defined in the reduced energy scale, relative to the potential surface of odd-odd nuclei.

---

[†] In the interest of clarity, a constant value of the pairing-gap parameter $\Delta$ is assumed. That means that local variations of $\Delta$ are neglected. The binding-energy surface of odd-mass fragments is assumed to be exactly in the middle between even-even and odd-odd fragments, meaning that neutron-proton residual interactions are also neglected. A more realistic description would not alter the main conclusions of this work.

The total level density is given by the product of the level densities of the two fragments $\rho_1(U_1)\rho_2(U_2)$. The most probable configuration is the one that provides the highest total level density. Starting from a situation where $U_1$ and $U_2$ are in the range of energy shown in Fig. 3, due to the very different logarithmic slopes of the level densities for the heavy and the light fragment group, the most favorable configurations are those that minimize the excitation energy of the light fragment. In other words, energy sorting takes place [1]. Only at the final stage of the energy sorting, when $U_1 \approx 0$ and $U_2 \approx U_{tot}$, the benefit of transferring the unpaired nucleons to the heavy fragment to form an even-odd, an odd-even or an even-even light fragment becomes apparent because only for these configurations there are states available. If an even-even light fragment is formed, instead of an odd-odd one, the energy in the heavy fragment increases to $U_2 = U_{tot} + 2\Delta_1$, which for the nuclei considered in Fig. 3 corresponds to an increase of the level density of the heavy fragment of more than three orders of magnitude, while $U_1 = -2\Delta_1$. It becomes clear that configurations of two fragments in contact where the light fragment is fully paired (i.e. it has no quasi-particle excitations) are strongly favored for very asymmetric fission.

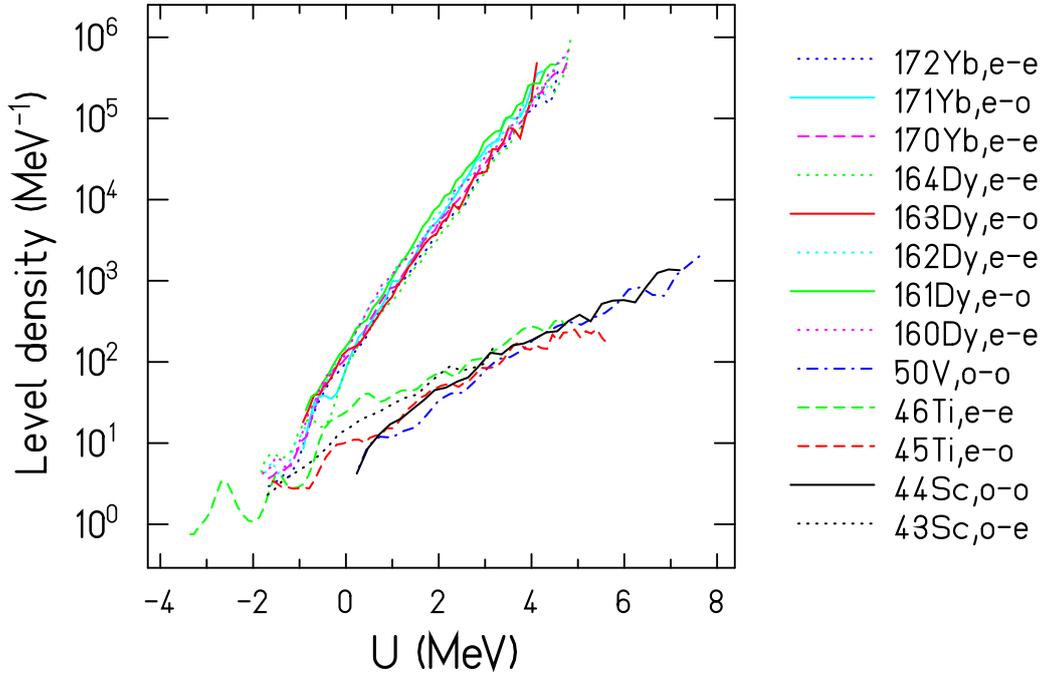

**Figure 3.** (Colour online) Experimental level densities of various nuclei [27-34] in a reduced excitation-energy scale $U = E_{gs} - n\Delta$. The excitation energy above the ground state $E_{gs}$ is reduced by $2\Delta$ ($n = 2$) for even-even (e-e) nuclei, by $\Delta$ ($n = 1$) for even-odd (e-o) or odd-even (o-e) nuclei and left unchanged ($n = 0$) for odd-odd (o-o) nuclei.

5. **Details of the model**

For an even-even fissioning nucleus, the number of configurations with $Z_1$ even is given by:

$$N^{ee}_{Z_1=e}(Z_1) = \int_{-2\Delta_1}^{U_{tot}+2\Delta_2} \rho_1(U_1)_{(ee)} \rho_2(U_{tot}-U_1)_{(ee)} dU_1 + \int_{-\Delta_1}^{U_{tot}+\Delta_2} \rho_1(U_1)_{(eo)} \rho_2(U_{tot}-U_1)_{(eo)} dU_1 \qquad (3)$$

where $\rho_i(U_i)_{(ee)}$ and $\rho_i(U_i)_{(eo)}$ are the level densities of representative even-even and even-odd nuclei, respectively, with mass close to $A_1$ or $A_2$. The number of configurations with $Z_1$ odd is:

$$N^{ee}_{Z_1=o}(Z_1) = \int_{-\Delta_1}^{U_{tot}+\Delta_2} \rho_1(U_1)_{(oe)}\rho_2(U_{tot}-U_1)_{(oe)}dU_1 + \int_{0}^{U_{tot}} \rho_1(U_1)_{(oo)}\rho_2(U_{tot}-U_1)_{(oo)}dU_1 \qquad (4)$$

where $\rho_i(U_i)_{(oe)}$ and $\rho_i(U_i)_{(oo)}$ are the level densities of representative odd-even and odd-odd nuclei, respectively, with mass close to $A_1$ or $A_2$. The level densities used in eqs. (3-4) are total level densities. Therefore, they include also the states with a fully paired proton subsystem in one or both of the fission fragments discussed in [15]. The integrals over the level densities $\rho_i(U_i)_{(ee)}$, $\rho_i(U_i)_{(eo)}$, $\rho_i(U_i)_{(oe)}$ and $\rho_i(U_i)_{(oo)}$ in eqs. (3-4) reflect that each unpaired nucleon can freely move from one fragment to the other. For instance, the first term in eq. (3) includes the configurations where the two nucleons of broken pairs remain in the same fragment and the second term the configurations where the neutrons of one broken pair are in different fragments while the protons of broken pairs are in the same fragment.

The yield for even-$Z_1$ nuclei is:

$$Y^{ee}_{Z_1=e}(Z_1) = \frac{N^{ee}_{Z_1=e}(Z_1)}{N^{ee}_{Z_1=e}(Z_1) + N^{ee}_{Z_1=o}(Z_1)} \qquad (5)$$

For an odd-even fissioning nucleus, we have:

$$N^{oe}_{Z_1=e}(Z_1) = \int_{-2\Delta_1}^{U_{tot}+\Delta_2} \rho_1(U_1)_{(ee)}\rho_2(U_{tot}-U_1)_{(oe)}dU_1 + \int_{-\Delta_1}^{U_{tot}} \rho_1(U_1)_{(eo)}\rho_2(U_{tot}-U_1)_{(oo)}dU_1 \qquad (6)$$

$$N^{oe}_{Z_1=o}(Z_1) = \int_{-\Delta_1}^{U_{tot}+2\Delta_2} \rho_1(U_1)_{(oe)}\rho_2(U_{tot}-U_1)_{(ee)}dU_1 + \int_{0}^{U_{tot}+\Delta_2} \rho_1(U_1)_{(oo)}\rho_2(U_{tot}-U_1)_{(eo)}dU_1 \qquad (7)$$

Similar equations hold for even-odd and odd-odd fissioning systems. In the reduced energy scale used in eqs. (3-4 and 6-7), the level densities of neighbouring even-even, odd-$A$ and odd-odd nuclei are very similar for positive reduced excitation energies (see Fig. 3). Therefore, the difference between the number of configurations $N_{Z_1=e}$ and $N_{Z_1=o}$ is essentially given by the integrals over $U_1 = -2\Delta_1$ to 0 and over $U_1 = U_{tot}$ to $U_{tot} + 2\Delta_2$. This shows that only the population of the energy states below the pairing gap of even-even nascent fragments (light or heavy) can cause an odd-even effect in fission. This happens when the excitation energy $U_{tot}$ and all the unpaired nucleons (protons and neutrons) are in the complementary fragment. At asymmetry, due to the higher level density of the heavy fragment, the main contribution to the odd-even effect comes from the configurations where the light fragment is fully paired.

For even-even fissioning nuclei, eqs. (3) and (4) show that $N^{ee}_{Z_1=e} > N^{ee}_{Z_1=o}$ for every charge split ($Z_1$ and $Z_2$) and the odd-even effect $\delta_p$ will always be positive, even at symmetry. In contrast, for odd-even fissioning nuclei $N_{Z_1=e}^{oe} \approx N_{Z_1=o}^{oe}$ close to symmetry because $\Delta_1 \approx \Delta_2$. $N_{Z_1=e}^{oe}$ starts to be larger than $N_{Z_1=o}^{oe}$ as we move to more asymmetric splits because $\Delta_1 > \Delta_2$. It is interesting to deepen into the origin of the local odd-even effect at symmetry according to our model. For even $Z_{CN}$ the number of unpaired protons is even. At asymmetry, excitation energy and all the unpaired protons are preferentially transferred to the heavy fragment. At symmetry the two fragments are very close in mass and excitation energy and unpaired protons may be transferred to either one or the other fragment. This will always lead to an increase of the yield of neighbouring even-Z fragments and to a positive $\delta_p$ (see eq. 1). On the other hand, for an odd-Z fissioning nucleus the number of unpaired protons is odd. At symmetry, the transfer of energy and all the unpaired protons to one of the fragments will increase both, the yield of neighbouring even-Z and odd-Z fragments. This is why for odd-Z fissioning nuclei $\delta_p$ is zero at symmetry. In fact, in our approach there is no difference in the nature of the odd-even effect at symmetry and asymmetry. In all cases the odd-even effect is caused by the preferential formation of a nascent even-even fragment with no quasiparticle excitations. In other words, there is a tendency for all the excitation energy and the unpaired nucleons to be found in one of the nascent fragments, at asymmetry this is the heavy fragment.

In practice one cannot always find experimental level densities representative of even-even, odd-A or odd-odd nuclei for each value of $Z_1$ and $Z_2$ covered by the fission yields. However, given the similarity between the experimental level densities of neighbouring nuclei in the reduced energy scale, we have replaced in eqs. (3-4 and 6-7) the representative level densities by the level densities $\rho_i$ of the two fission fragments considered, namely $A_1$, $Z_1$ and $A_2$, $Z_2$. The level densities $\rho_i$ are obtained using a composite formula with a constant-temperature description at low excitation energies and a shifted Fermi-gas description above. In order to include the effects of pairing correlations on the nuclear level density in a consistent way, the recommended energy shift of the Fermi-gas part of the Gilbert-Cameron composite formula [35] is increased by 2 MeV [36]. Consequently, in our formula the transition from the constant-temperature to the Fermi-gas regime occurs at excitation energies $E_{gs}$ of about 8-9 MeV which are higher than the transition energies of the broadly used Gilbert-Cameron formula.

Fig. 4 shows the level densities of various nuclei used in our calculations and obtained with the composite level-density formula described above. As can be seen, for less asymmetric splits, the logarithmic slopes of the level densities of the two fragments get closer. Therefore, the probability to populate the lowest-energy states of even-even or even-odd emerging light fragments decreases, leading to a decrease of $\delta_p$ with decreasing asymmetry. Fig. 4 also illustrates that the logarithmic slope of the Fermi-gas part of the level density gradually decreases with increasing excitation energy. This implies that the relative statistical weight of configurations with a fully-paired light fragment will be less important than in the constant-temperature regime. Thus, the transition from the constant-temperature to the Fermi-gas regime that may occur when $U_{tot}$ increases will lead to a considerable decrease of $\delta_p$. This explains why $\delta_p$ decreases with increasing mass of the fissioning nucleus.

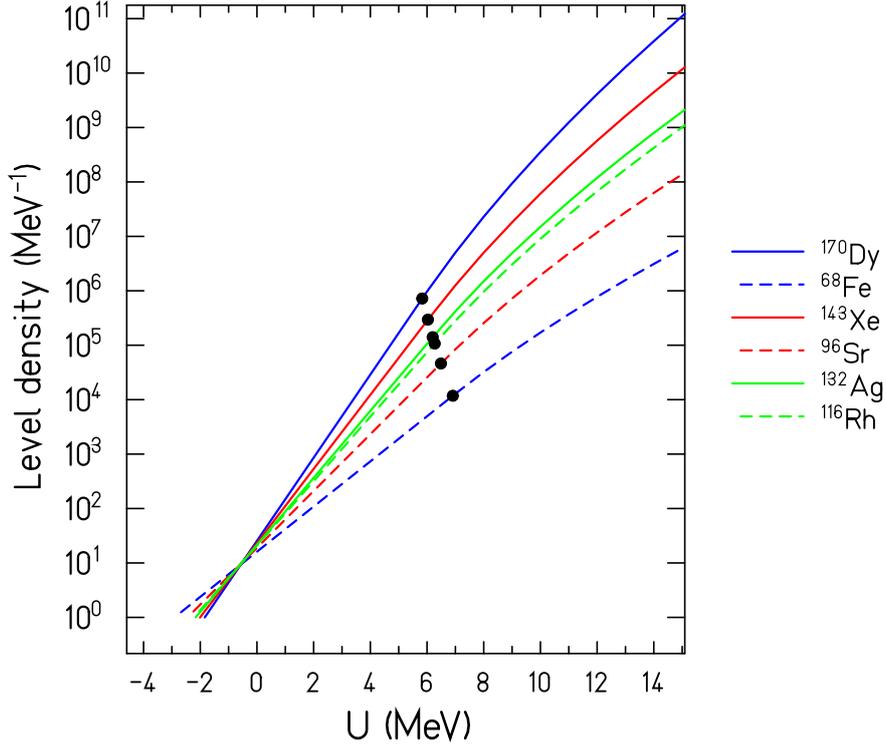

**Figure 4:** (Colour online) Level densities in an energy scale $E_{gs}$-2$\Delta$ with $\Delta = 12/\sqrt{A}$. The curves were obtained using the composite level-density formula of ref. [36]. The level densities of complementary fission fragments are represented with the same colour. The full dots indicate the excitation energies where the transition from the constant-temperature to the Fermi-gas regime occurs.

As said above, $U_{tot}$ increases with the dissipated energy. We obtain best agreement with fission data (including a wide variety of fission observables and fissioning nuclei [37]) when we consider that 35% of the potential energy difference from saddle to scission [20] is dissipated. According to this, the dissipated energy varies from about 3 MeV for $^{230}$Th to 8.5 MeV for $^{250}$Cf.

### 6. Level densities of nascent fission fragments

In this section we assess the adequacy of the level-density formulae described above in a situation before scission where the touching fragments are strongly deformed due to their mutual Coulomb repulsion.

From a macroscopic point of view, the use of ground-state level-densities at significantly larger deformations is well justified. In fact, according to e.g. ref. [38] the level-density parameter *a* of the Fermi-gas level density at a deformation as large as $\beta$=0.6 is only about 1 % larger than in the ground-state deformation. In [39] it was shown that there is a strong anticorrelation between the level-density parameter *a* and the inverse logarithmic slope of the constant-temperature formula, the so called temperature parameter *T*. Therefore, it follows that the temperature parameter of the constant-temperature level density at very large deformations is less than 1% smaller than at the ground state.

In this work, we have not included the influence of shell effects on the level density. Egidy et al. [40] obtained the following empirical dependence between the temperature parameter $T$, the mass of the nucleus $A$, and shell effects $S$:

$$T = \left(\frac{d\ln(\rho^{CT}(E))}{dE}\right)^{-1} = \frac{1}{A^{2/3}}(17.5 - 0.51S + 0.051S^2) \qquad (8)$$

Low-energy fission of actinides is governed by deformed shells in the heavy fragment that are significantly weaker than major shells at the ground state. Indeed, the strongest contribution to the fission yields of heavy actinides is given by the asymmetric S2 fission channel which is caused by a deformed shell located around $A = 140$. This shell has a depth of about 4 MeV [37]. Setting $S = -4$ MeV in eq. (8) leads to an increase of $T$ by about 16%. Considering the influence of mass asymmetry, for a split into $A_1 = 140$ and $A_2 = 96$ in the fission of $^{236}$U, the light fragment would still have a temperature parameter that is about 13% higher than the one of the complementary heavy fragment. Thus, energy sorting would only be weakened, but not reversed. The spherical shell closure around $^{132}$Sn is stronger. Its maximum value is $S \approx -7.4$ MeV [37] when $^{132}$Sn is formed. Note that the magnitude with which the effect of the shell closure manifests depends on the relative yield of the different fission channels. The contribution to the yields of the S1 channel for heavy actinides is much weaker than the one of the S2 channel. The S2 channel represents about 99% of the total yield for the thermal neutron-induced fission of $^{235}$U, for example. Moreover, the maximum shell effect behind the S1 fission is very much localized around the doubly magic $^{132}$Sn. The shell is much weaker for the more neutron-deficient fragments that are predominantly produced in the fission of the most commonly available actinides [37]. In addition, there might be another shell in the complementary light fragment‡ that would increase the temperature of the light fragment and thus counteract the influence of the shell in the heavy fragment on the energy-sorting process. Nevertheless, an eventual inversion of the energy-sorting process in the region of the S1 fission channel cannot be excluded. However, there is no clear indication for such an inversion from the variation of the mass-dependent prompt-neutron yields with incident neutron energy measured in [41]: The prompt-neutron yield of the light fragments remains constant within the experimental uncertainties. An observation of an eventual inversion of the energy-sorting process by the odd-even effect for even-Z fissioning systems is not possible, because the even-odd effect is insensitive to the sign of the temperature difference. This is not the case for odd-Z fissioning systems, but any data that would show such an effect are not available. Due to the described complexity of the problem and the lack of relevant data, the influence of shell effects on the energy-sorting process is not considered in this work.

In our description, we also assume that the level densities follow the constant-temperature behaviour down to $U = -\Delta$ for odd-$A$ and down to $U = -2\Delta$ for even-even nuclei. As shown in

---

‡The strong mass dependence of the prompt-neutron yields in the light-fragment region gives an indication for a sizeable influence of shell structure also in the light fragment.

Fig. 3, this is rather reasonable for most of the fission fragments. The reason is the presence of collective levels and the energy resolution of the measurements. In addition, the observed irregularities in the level densities are considerably smoothed by the thermal averaging caused by the strong fluctuations of the energy exchange [2].

## 7. Results and discussion

The results of our calculation are compared with experimental data in Fig. 5. The increase of $\delta_p$ with asymmetry and the decrease of $\delta_p$ with increasing mass of the fissioning nucleus are fairly well reproduced. Only for $^{230}$Th the measured values are substantially underestimated by the model. For $^{236}$U, the data point that is closest to symmetry is appreciably higher than the calculation. This effect may be associated to the influence of the $Z$=50 shell in the complementary fragment, which is known to enhance the yield of tin isotopes. Similarly, the most asymmetric data point of $^{246}$Cm is exceptionally high, which may be due to the influence of the $Z$=28 shell. Our calculation is also in good agreement with the data for the odd-Z fissioning nucleus $^{243}$Am, which shows an odd-even effect of similar magnitude as even-Z fissioning nuclei of comparable mass. This is particularly interesting and clearly demonstrates the strong influence of complete energy sorting (i.e. the formation of an even-even light fragment with no quasiparticle excitations) on the odd-even effect in fission. Indeed, an important contribution to the odd-even effect from heavy fragments with a fully paired proton configuration would enhance the odd-even effect for even-Z fissioning systems and reduce the odd-even effect for odd-Z fissioning systems, leading to an odd-even effect of different magnitude for the two types of fissioning nuclei. Our calculation gives a zero local odd-even effect at symmetry for $^{243}$Am. There are no experimental data at symmetry for this nucleus, but the electromagnetic-induced data measured in inverse kinematics at GSI confirm that there is no local odd-even effect at symmetry for odd-Z fissioning nuclei [14].

When $U_{tot}$ is small, as is the case for $^{230}$Th, $\delta_p$ varies very rapidly with $U_{tot}$. Therefore, $^{230}$Th is particularly sensitive to the uncertainties on the dissipated energy. The disagreement found for $^{230}$Th may be caused by the neglect of fluctuations in the dissipated energy. In fact, for a great part of the fission events the available energy may be so low that they reach the scission point in a completely paired configuration due to the threshold character of the first quasi-particle excitation.

The time to form a fully-paired light nascent fragment is the sum of the time needed for the light fragment to transfer its energy to the heavier one and the time to transfer few unpaired nucleons through the neck. If this time is longer than the saddle-to-scission time, our model will over predict the magnitude of the odd-even effect. Therefore, the general agreement between the experimental data and our calculation suggests that the time for the population of a fully-paired light fragment in accordance with statistical equilibrium is shorter than the saddle-to-scission time. It would be interesting to investigate whether microscopic models can confirm this finding.

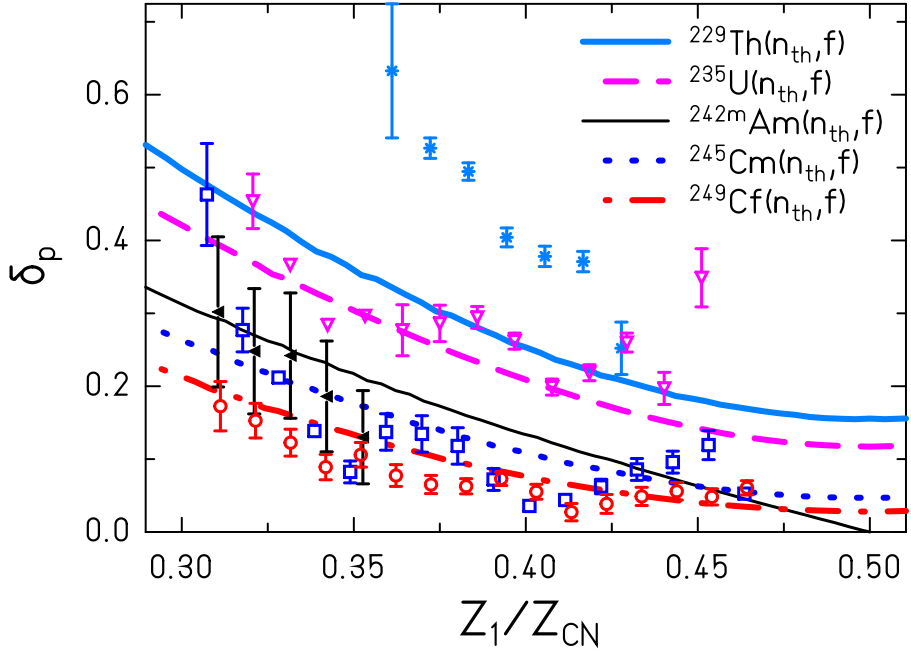

**Figure 5:** (Colour online) Local even-odd effect $\delta_p$ as a function of asymmetry. The symbols represent experimental data from the compilation of [7] and denote the target nuclei: $^{229}$Th (stars), $^{235}$U (open triangles), $^{242}$Am (full triangles), $^{245}$Cm (open squares), $^{249}$Cf (open circles). The lines correspond to the results of the model developed in this work.

During neck rupture there may be additional pair breaking [10, 11]. However, there is not enough time at scission to distribute this additional excitation energy and unpaired nucleons according to statistical equilibrium. The general good agreement between our calculations and the data suggests that the unpaired protons generated at scission stay confined in the fragments. Indeed, if we would include in our model the possibility that each proton of one of these additional broken pairs ends in a different fragment, the odd-even effect would be reduced. Our results also indicate that the variation of the excitation energy with asymmetry discussed in [16] should be rather moderate for most of the fissioning systems we have considered. This effect would however lead to a better agreement with the data for $^{230}$Th, as reducing the excitation energy for asymmetric splits would increase the odd-even effect. Also the influence of shell effects that are neglected in our calculations seems to be small.

We would like to stress that our approach differs substantially from previous attempts [10] to explain the odd-even effect in fission-fragment Z distributions by a Boltzmann-like factor $\exp(\Delta Q/T)$, where $\Delta Q$ is the systematic difference in the $Q$ values of different classes of charge splits (e.g. even-Z – even-Z and odd-Z – odd-Z for even-Z fissioning nuclei) and $T$ is a suitable temperature parameter. Due to the fundamental difference of a Boltzmann gas and the nucleonic Fermi gas, eventually under the influence of pairing correlations, it is essential for a realistic application of the statistical model to nuclei to consider the specific features of the nuclear level density. Note that systematic odd-even fluctuations in the nuclear level density do essentially not exist on an absolute energy scale, represented by $U$ in the present work, above the binding-energy surface of odd-odd nuclei (see also [14], Fig. 3 of [42] and Fig. 4 of [43]). Systematic differences of the level densities are only created by the presence of additional levels in odd-mass and even-even nuclei below the binding energy surface of odd-odd nuclei.

## 8. Conclusion

We have used statistical mechanics to investigate the equilibration of a fissioning nucleus from the point where the properties of the nascent fission fragments are well defined. The odd-even effect of fission-fragment yields has been calculated in the most direct and complete way by comparing the number of available final states for the formation of a light fragment with an even number of protons to those of a neighboring fragment with an odd number of protons. The number of final states has been calculated with nuclear level densities in which the effects of pairing correlations are consistently included. The smooth variation of the number of states as a function of mass asymmetry has been corrected for by using a reduced excitation-energy scale. We have considered a given fixed excitation energy. The inclusion of an excitation-energy distribution and a possible variation of the available excitation energy with the asymmetry of the mass split discussed in [16] can easily be done if the necessary information is available from some theoretical estimation. However, the general good agreement found between our calculations and most of the experimental data indicate that these additional effects might only have a weak impact.

Our model shows that the general dependence of the odd-even effect in fission-fragment yields with the mass of the fissioning system and with the mass asymmetry of the fragments can be explained by the higher statistical weight of configurations in which the light nascent fragment is an even-even nucleus with no quasi-particle excitations. This implies that excitation energy and unpaired nucleons are predominantly transferred to the heavy fragment before scission. Thus, within our model, the odd-even effect in fission-fragment yields represents a second strong evidence for the importance of the entropy-driven energy-sorting mechanism, which is fully consistent with the excitation-energy dependence of the prompt-fission neutron yields [1]. Whereas prompt-neutron yields demonstrate how the energy sorting determines the average excitation-energy of the two fragments, the odd-even effect reflects complete energy sorting, that is, the population of the ground state of the light fragment.

It would be desirable to further verify the validity of our model with additional high-quality experimental data on the odd-even staggering and the mass-dependent prompt neutron yields. In particular, the measurement of new data spanning a broad range of asymmetry at different initial excitation energies and for a wide variety of fissioning nuclei is highly encouraged.

**Acknowledgements**

This work was supported by the European Commission within the Sixth Framework Programme through EFNUDAT (project no. 036434) and within the Seventh Framework Programme through Fission-2010-ERINDA (project no.269499).